\documentclass[10pt]{article}

\usepackage{amsmath}
\usepackage{amssymb}

\usepackage{graphicx}

\usepackage{cite}

\usepackage{color} 

\usepackage{setspace} 
\usepackage[labelfont=bf,labelsep=period,justification=raggedright]{caption}

\makeatletter
\renewcommand{\@biblabel}[1]{\quad#1.}
\makeatother

\date{}

\usepackage{url}

\begin{document}

\begin{flushleft}
{\Large
\textbf{Low-bandwidth and non-compute intensive remote identification of microbes from raw sequencing reads.}
}
\\
Laurent Gautier$^{1, 2, \ast}$, 
Ole Lund$^{1}$
\\
\bf{1} Center for Biological Sequence Analysis (CBS),
Department of Systems Biology, Technical University of Denmark,
Lyngby, Denmark
\\
\bf{2} DTU Multiassay Core (DMAC),
Technical University of Denmark,
Lyngby, Denmark
\\
$\ast$ E-mail: laurent@cbs.dtu.dk
\end{flushleft}

\section*{Abstract}

Cheap high-throughput DNA sequencing may soon
become routine not only for human genomes but also for practically anything
requiring the identification of living organisms from their DNA:
tracking of infectious agents, control of food products, bioreactors, or
environmental samples.

We propose a novel general approach to the analysis of sequencing data
in which the reference genome does not have to be specified.
Using a distributed architecture we are able to query a remote server
for hints about what the reference might be, transferring a relatively
small amount of data, and the hints can be used for more
computationally-demanding work.

Our system consists of a server with known reference DNA
indexed,
and a client with raw sequencing reads. The client sends a sample
of unidentified reads, and in return receives a list of matching references
known to the server. Sequences for the references can
be retrieved and used for exhaustive computation on the reads, such
as alignment.

To demonstrate this approach we have implemented a web server, 
indexing tens of thousands of publicly available genomes and genomic regions
from various organisms and returning lists of matching hits from query
sequencing reads.

We have also implemented two clients, one of them running in a web browser,
in order to demonstrate that gigabytes of raw sequencing reads
of unknown origin could be identified without the need to transfer 
a very large volume of data, and on modestly powered computing devices.
The browser-based client can run from a tablet, perform its task
within seconds, and consume an amount of bandwidth that would make the
use of a mobile broadband network practical. 
We anticipate that such client-server approaches will develop in the future,
allowing a fully automated processing of sequencing data without the need
to specify what the reference is.
It is already in use at our core facility for routine instant quality check of
sequencing runs from desktop sequencers.

A web access is available at \url{http://tapir.cbs.dtu.dk}.
The source code for a python command-line client,
a server, and supplementary data are available at \url{http://bit.ly/1aURxkc}.


\section*{Introduction}
The sequencing of DNA has become increasingly affordable during the last
decade\cite{Shendure_Next-Generation_2008} and modern high-end sequencers
have the capacity to process the
equivalent of several human genomes or several hundred bacteria.
Current desktop sequencers require limited initial investments,
and are providing flexibility over sequencing volumes.
The sequencing of complete bacterial genomes from isolates can
be performed in a day.
Recent announcements on nanopore sequencing\cite{Rusk_Cheap_2009}
are even suggesting that sequencers could be so cheap that they would
be disposable.
Extracting DNA is itself a relatively simple procedure, and it is foreseeable
that DNA sequencing will soon be a relatively cheap routine procedure in
molecular biology. Patients will be sequenced routinely, outbreaks
of infectious agents traced by their DNA, quality of water and
food also monitored with DNA sequencing.

On the analytic side, local alignment of sequences, with pioneering tools such as
the Smith-Waterman algorithm\cite{Smith_Identification_1981}, has been a cornerstone of bioinformatics.
Once applied between a query and a collection of references it allowed the scoring and ranking of alignments,
letting researchers
infer the origin and function of newly sequenced DNA or RNA from its similarity to already
existing sequences. Although the methodology has come under criticism for being inaccurate
at times\cite{Devos_Practical_2000, Rost_Enzyme_2002}, its popularity remains indisputable with a large number of
functional annotations in public databases having the mention `by sequence homology'.
However, aligning newly obtained DNA to existing references archived in database remains
a relatively demanding computational task. BLAST\cite{Altschul_Basic_1990} and 
later BLAT\cite{Kent_Blat---Blast-Like_2002} increased
the speed at which alignments can be done, although at the cost accuracy,
yet with the number of sequences currently available searching a sequence
against the pool of known sequences still requires significant compute ressources, and despite
all its computing power the NCBI do not recommend submitting to their servers more than 50 query sequences at once.
New tools designed for short-read
sequencing have since been developed, such as Bowtie\cite{langmead_ultrafast_2009},
BWA\cite{Li_Fast_2009}, and SHRiMP\cite{rumble_shrimp_2009} to name only a few,
but these tools are designed to align all sequencing
reads against a given known reference. Li {\it et al}\cite{li_survey_2010} can be consulted for
a more comprehensive review of short-read aligners.
In order to achieve speed, such tools load an index of the reference into memory, and thus
limit the amount of reference DNA that can be handled unless costly computing infrastructure with
a lot of shared memory is used.

\section*{Results}

  \subsection*{Matching a sample of sequencing reads against a comprehensive collection of references}

A subjective way of looking at the alignment programs is to split them
into two main categories: the ones designed to map one query sequence
against a collection of known reference (e.g., BLAST), and the ones designed
to map a large number of short sequences against one specified reference
as quickly as possible (e.g., bowtie or BWA). We propose an intermediate
approach in which candidate references are quickly identified before more
detailed work, such as aligning the reads, is performed.
We use a small random sample of initial data to query a server acting
as a \textit{DNA search engine}, thus
consuming a small amount of bandwidth when compared to approaches that
transfer all sequencing data to computation servers.
Using synthetic reads generated from full genomes, we demonstrate that identifying
a pure culture from raw DNA sequencing reads can be achieved with 100 random
reads.

When building the database of references, we do not require a feature
selection step and the building of a classifier. We simply index
{\it k}-mers found in the reference genomes,  simplifying greatly the complexity of the procedure.
This comes at the cost of space usage, 
with potentially less informative {\it k}-mers being 
indexed, but also comes with the advantage of being linear
in the total size for the collection of references and relatively easy to
parallelize.
This makes the indexing of all known DNA, like
web search engines index all documents on the internet, a plausible
eventuality.

Our approach also differs from alignment algorithms, in which one sequence
is matched against a collection of reference sequences, as we consider several
sequences (sequencing reads) against a collection of reference sequences, and vote which
references are most representative of the query set. Each read sent
to the server is matched against the reference sequences, 
and a summary of candidate DNA references over all the reads is returned.

\begin{figure}[!ht]
\begin{center}
\includegraphics[width=.8\textwidth]{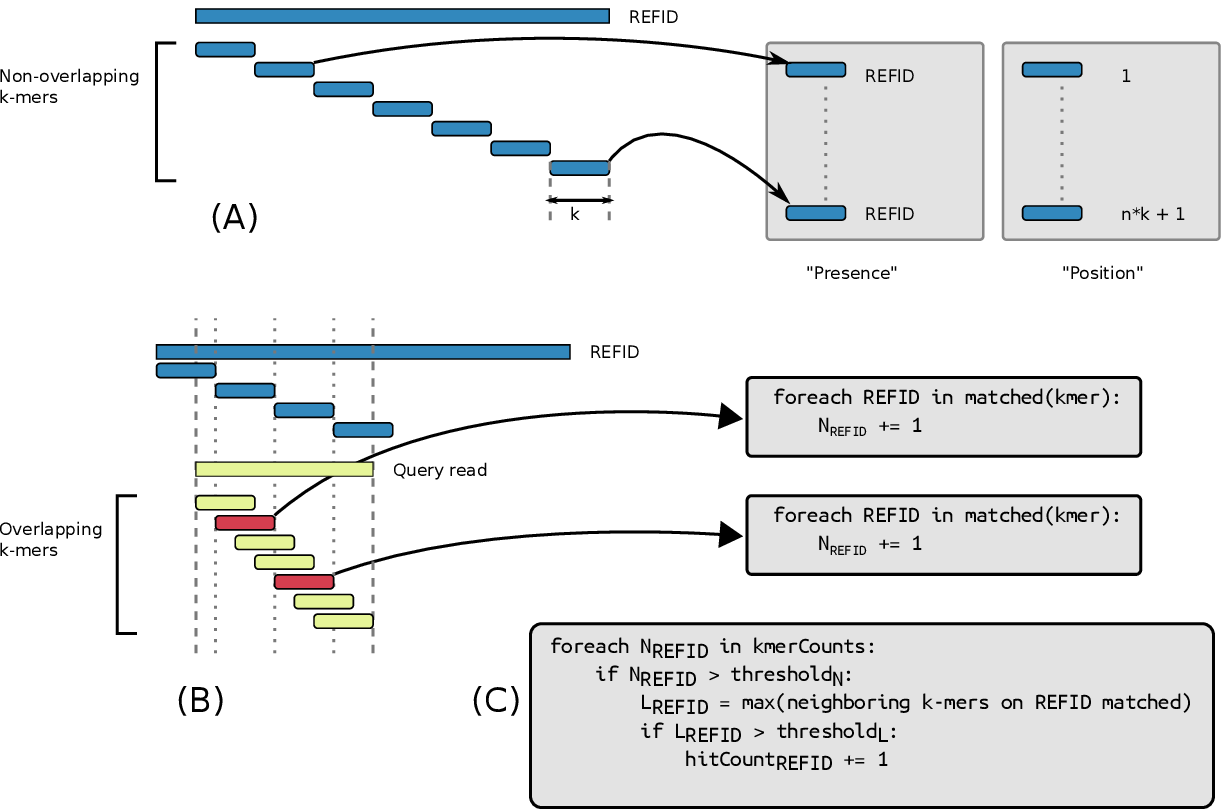}
\end{center}
\caption{{\bf Overview of the indexing and scoring procedures.}
(A) During the indexing of a collection of reference sequences,
non-overlapping {\it k}-mers are indexed into two distinct key-value stores, one associating
{\it k}-mers with the references they were found in (`presence') and one
associating {\it k}-mers with the position in the reference at which the k-mer
was found (`position'). (B) When processing a sequencing read in a query
set, overlapping {\it k}-mers are looked up in the `presence' store. Using overlapping
{\it k}-mers allows to resolve relatively rapidly misalignments between the
beginning of the read and the beginning of the reference sequence (dotted lines). On
the figure, only {\it k}-mers in red are in phase with the indexing step, 
therefore only those can be found in `presence'. (C) For a given read, a threshold
is applied to retain only references potentially matching enough of the read.
Situations where very large references containing disjoint scattered {\it k}-mers,
such as a bacterial read against a mammalian genome, are resolved in the last step
where the `position' store is queried.
}\label{fig:overview}
\end{figure}

The matching of reads is done by splitting them into overlapping {\it k}-mers and
searching whether {\it k}-mers are associated with a reference. {\it k}-mers
are not disssociated from the sequencing read they come from and we are looking for clusters of matching {\it k}-mers,
close to one another although not necessarily in the same order on both
the query read and the reference sequence, as shown in Figure~\ref{fig:overview}.

We currently use non-overlapping {\it k}-mers for the indexing while
we use overlapping {\it k}-mers in the queries, but we consider this
an implementation detail. We could easily use overlapping {\it k}-mers
for the indexing and non-overlapping {\it k}-mers in the queries while
keeping the same guiding principles for giving scores to matching references.
The trade-off is that non-overlapping {\it k}-mers in the database take less space,
but will necessitate more queries since the {\it k}-mers in the reads will have to be
overlapping.

When having a query set of DNA data to identify, such as raw reads from a sequencer
for the purpose of diagnostics,
we consider the brute-force approach that consists in mapping all reads against comprehensive
reference databases to have two main disadvantages: hundreds of megabytes or gigabytes
of data must be transferred from the sequencing facility to a computing center,
and the computing resources necessary to perform the task are significant.
Assuming that a reference collection contains 10,000 {\it E.coli}-sized bacteria and that it
takes 30 seconds for an optimized aligner such as BWA or bowtie2 to process 250 Mbases of raw
sequencing data (about 60x in average coverage if the genome is 4 Mbases in size), it would
take 3.5 days on a CPU, although this could be parallelized trivially on multiple CPUs.
Refinements such as the the concatenation of the genomes, or the use of a generalized suffix arrays
can be made but at the cost of requiring
ever increasing amounts of memory for a given monilithic index, particularly for the suffix and
post-processing computation to assign mapping positions to initial
reference genomes, and inevitable multiple matches as close genomes are
referenced, something that short read aligners do not report by default\cite{hoffmann2009fast}.
The time complexity of
locating the $n$ occurrences of a string of length $p$
in a reference of size $u$ using an FM-Index has an upper bound of $O(p + n \log^\epsilon u)$\cite{ferragina2000opportunistic},
meaning that although the complexity grows slowly as the size of the reference
increases, with a term in $log^\epsilon u$ ($0<\epsilon<1$),
it grows linearly with the number of highly
similar genomes (leading to multiple occurences).
Of noteworthy interest is the similarity in time complexity for locating the $n$ occurrences
of a string of length $p$ in a B-tree indexing non-overlapping 
{\it k}-mers in $u$ as we have implemented it. The time complexity for searching one of the $\frac{u}{k}$
non-overlapping
{\it k}-mers indexed is $O(\log\frac{u}{k})$\cite{cormen2001introduction}, to which iterating through already sorted $n$ matching positions
must be added: $O(\log\frac{u}{k} + n)$. This is repeated for the $p-k$ overlapping {\it k}-mers
in the query sequence: $O((p-k)(\log\frac{u}{k} + n))$. Time complexity
for exact matches is clearly in favour of the FM-Index, but the sampling keeps the number of strings
to look up rather low and the existing B-tree in NoSQL solutions allowed us to quickly implement a working prototype. Furthermore, when one considers that we do not have one reference
of size $u$ but $r$ references of average size $\frac{u}{r}$, and that we need to keep track of which
reference is matching, the naive
approach evoked earlier becomes $O(r(p + n \log^\epsilon \frac{u}{r}))$ (each reference matched in turn)
while an approach using concatenation of the references and a {\it post hoc} assignment of the matches
to the corresponding reference sequences using a binary search would run in
$O(n\log r + (p + n\log^\epsilon \frac{u}{r}))$, or $O(p + n (\log r + \log^\epsilon \frac{u}{r}))$.
Burrow-Wheeler based mapping algorithms have largely focused on putting data in memory to achieve speed,
historically with the goal of having a human genome fit in the RAM of cluster nodes prevalent
at the time\cite{langmead_ultrafast_2009}, although more recent efforts have looked at applying
them to increasingly large collections of sequences\cite{cox2012large},
while our approach embraces the perspective
of enormous reference databases and does not try to keep it all within the RAM of one computer.

\begin{figure}[!ht]
  \begin{center}
\includegraphics[width=.4\textwidth]{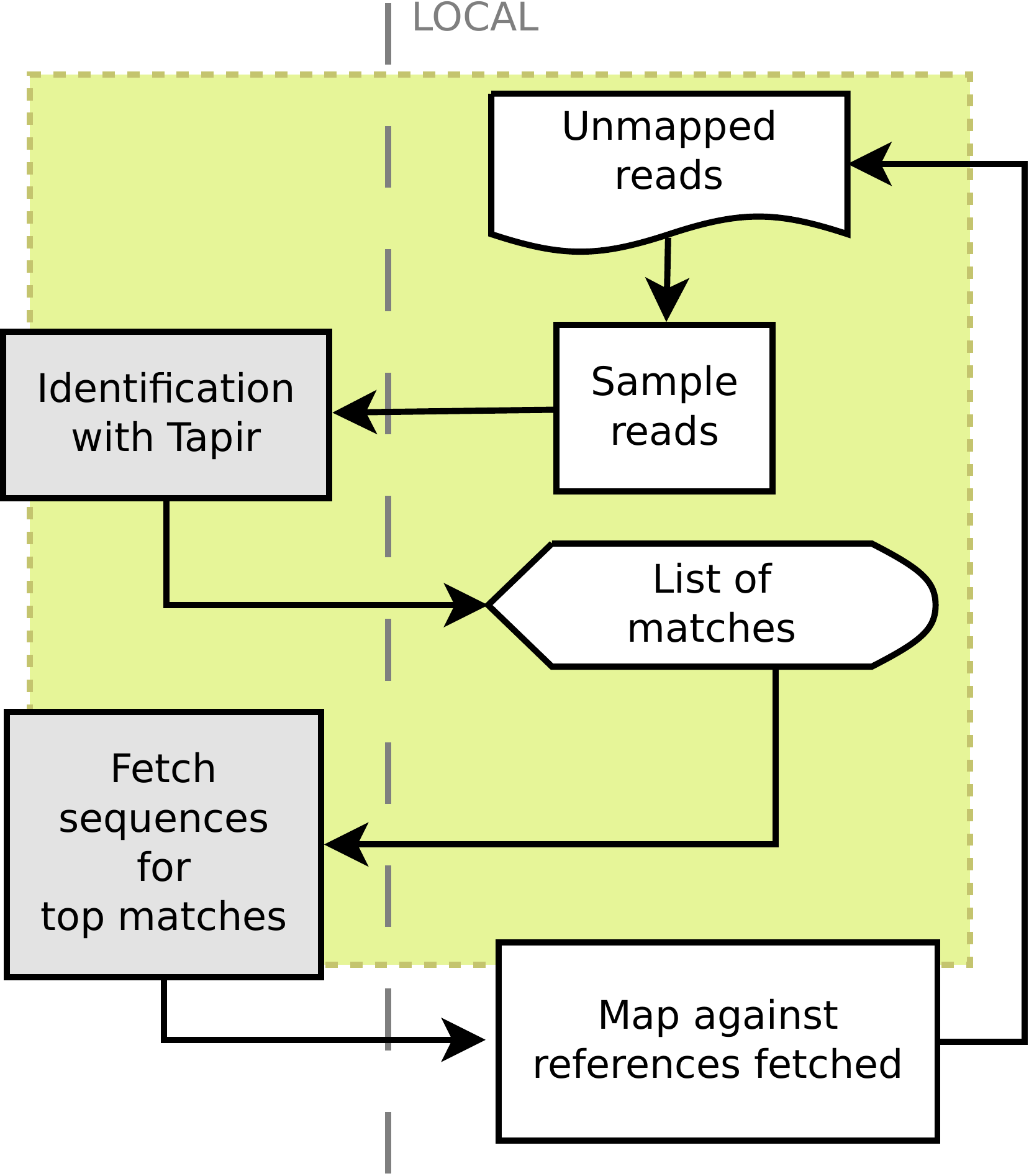}
  \end{center}
  \caption{{\bf Workflow.} Quick identification of DNA as performed with Tapir is a step
    in an analysis workflow when working with unknown samples. The current web browser-based
    client implements the part of the process in the grey area,
with the downloading of reference sequence currently in test and made available on
our production server very soon.
At the beginning, all reads are unmapped and a sample of them is submitted for identification.
The resulting list contains a pointer to the reference DNA represented most in the sample,
and the sequences for the top hits can then be fetched, indexed and used for mapping
all unmapped reads, for example with an aligner for short reads.
If unmapped reads remain after this step, they constitute a new set of unmapped reads
to iterate on. This procedure works by iteratively decreasing the number of reads; should
a mixture of DNA such as plasmids, or different species be present they will remain as 
unmapped and be handled with the next iteration.}
\label{fig:workflow}
\end{figure}

In addition to time complexity, the data transfer would be 250 Mbases of DNA,
with the sequencing data
moved to a data center that holds the references.
Our approach tries to restrict the search space for detailed investigation such as aligning all reads, or
SNP calling, or even template-based {\it de-novo} assembly, to a small set of references;
when evaluating performances we arbitrarily chose to initially consider a search a success
only if the right answer
is within a set of 5 proposed matches. The task of mapping all reads against
these references in order to identify precisely the best match
can be performed in 12 minutes on the same CPU, or in much less if a powerful multicore
architecture.
Transferring all genomes would represent about 20 Mbases of DNA, which could be performed
even over a 3G mobile internet connection. Our approach would enable a mobile sequencing facility
such as the Ion bus\cite{ionbus} to perform critical diagnostics or scientific tasks in remote
locations on the field.
Should there be unmapped reads, because of the presence of smaller DNA sources such as a plasmid,
virulence genes, a virus, or a mixture of bacteria, these reads can be processed similarly
and the full content be identified over a few iterations (see Figure~\ref{fig:workflow}).

Alternatives to brute force mapping require the building of a classifier,
which becomes an increasingly computationally demanding problem as the
collection of references increases.

  \subsection*{Building a benchmark}

To benchmark our system, originally designed to identify bacteria in sequencing data,
we took what was all sequences from bacteria available from the EBI
circa the beginning of 2012, that is 747 bacterial genomes while the full database of references
contained in addition to those: bacterial references from the NCBI, phages and viruses, plasmids,
and the human genome (see Table~1). For
each genome, we generated random possibly overlapping sub-sequences from the
genome sequence in order to simulate reads obtained from a DNA sequencer;
sub-sequences of length 50, 100, 150, 200, and 250 bases were used. We also
introduced uniform random substitutions of bases with rates of 0\% (no error), 1\%, 5\%, and 10\%
in order to simulate both a class of sequencing errors and the presence of point mutations
in real samples. For each genome, length, and substitution rates, a random sample of
100 sub-sequences, or reads, was performed and that sampling repeated 3 times.

We use synthetic reads from genomes we referenced, rather than a leave-one-out validation,
because we place ourselves in
a context of information retrieval, not machine learning and classification. Given the pace
at which new organisms are sequenced our hypothesis is that a building a central index containing
most of the DNA found in organisms will be possible. With the approach we present here,
sequencing sites would simply query that central server and only retrieve the relevant reference
genomes to perform detailed on all the reads locally. The volume of data transfer required is minimal,
making it practical for sequencing facilities in the field, and the use of central reference would
also allow make the update of the collection of reference shared across all users of the service.

  \subsection*{Prediction performances}
\begin{figure*}[!ht]
\begin{center}
\includegraphics[width=.8\textwidth]{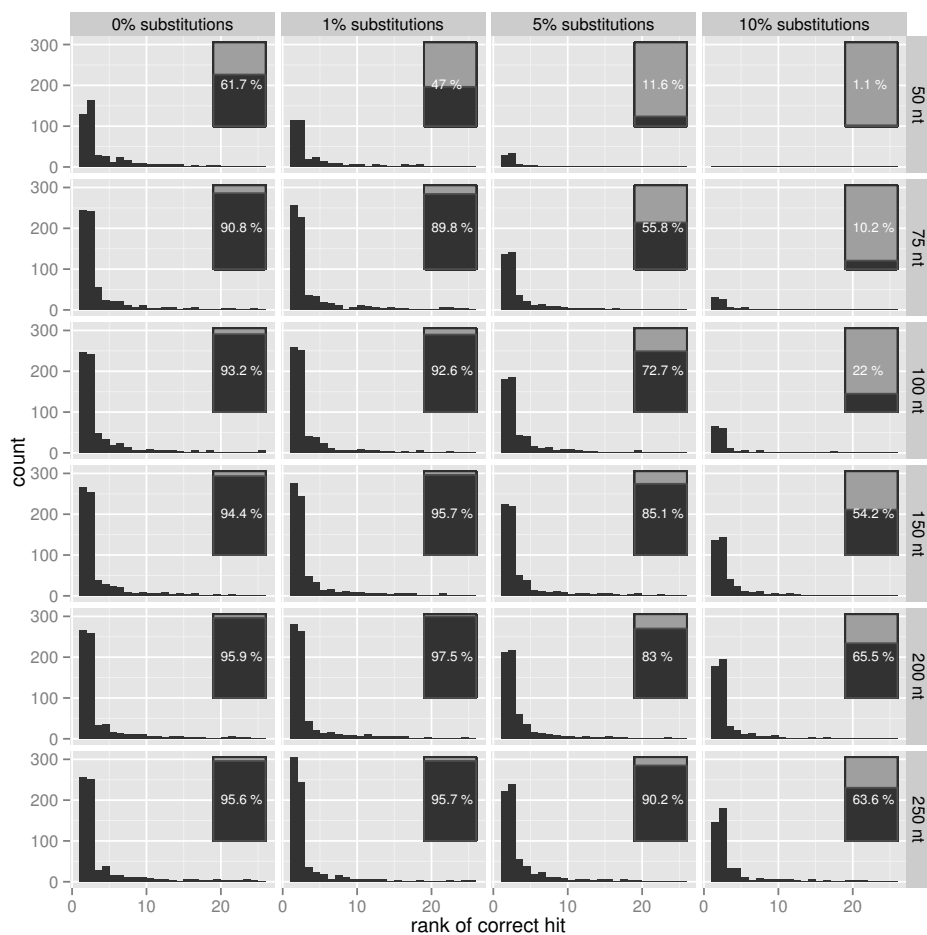}
\end{center}
\caption{{\bf Bacterial reads.} 
For each bacterial genome in a set of 747 genomes, we simulated several read lengths
(50 nt, 75 nt, 100 nt, 150 nt, 200 nt, 250 nt) and several substitution error rates
(0\%, 1\%, 5\%, 10\%). 100 random reads were used in each query and the distribution
of the rank of the correct references in the list recorded; a rank of $1$ means that
the correct reference was at the very top of the list. 
The list of hits has a maximum length of 25 and we count the reference as `not found' if
not in the list at all. The percentage of correct test bacterial genomes present in the list
is represented
in a bar nested on the right side of each panel. The figure shows that, as expected, the
performances degrade as the substitution rate increases, but also that reads of length 50
appear of little practical use for identification purposes. Increasing the read
length beyond 100 nt brings only small improvements, and has a limited
compensatory effect on the substitution rate. The figure suggests that current leading technology
for sequencing possess sufficient length for an accurate identification, and should
focus on sequence quality rather than increased read length.
}
\label{fig:perf}
\end{figure*}

\begin{figure*}[!ht]
\begin{center}
\includegraphics[width=.9\textwidth]{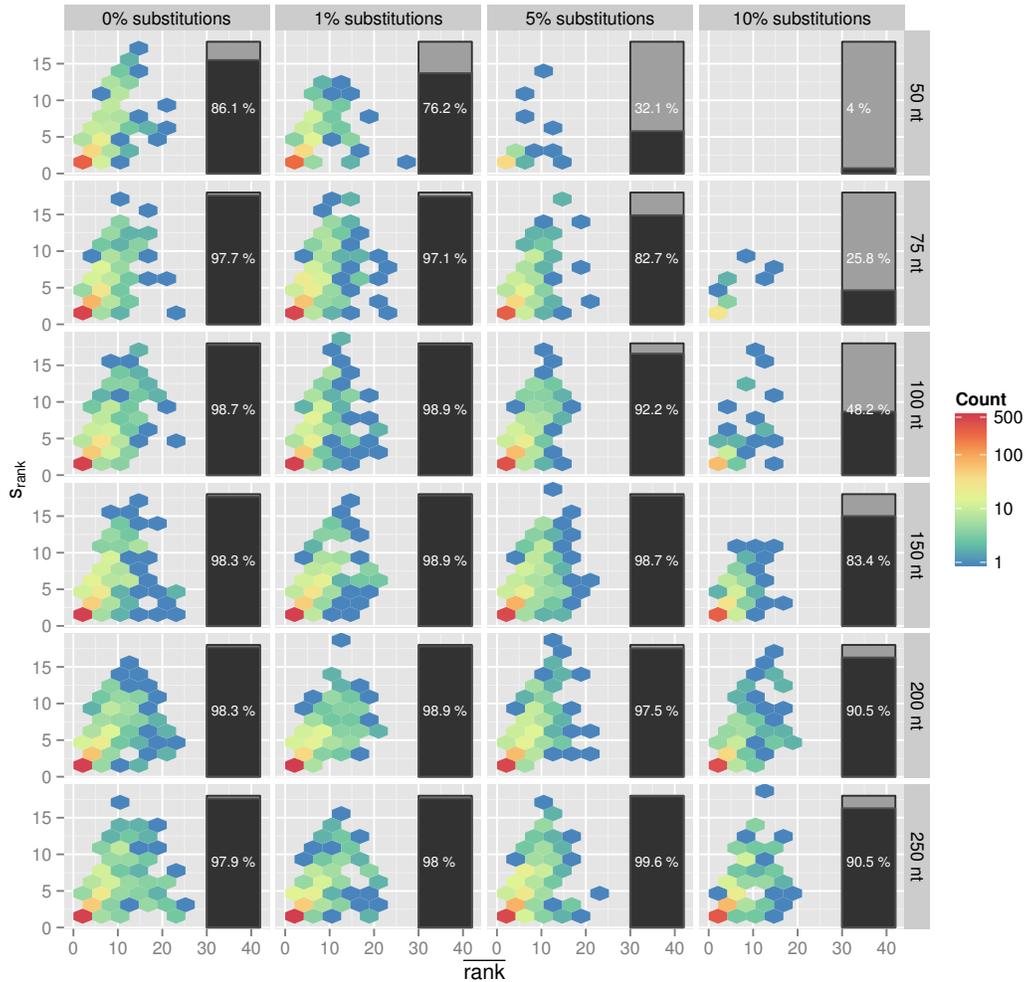}
\end{center}
\caption{{\bf Bacterial reads, variability of performances} 
Average rank ($\overline{\mathtt{rank}}$, x-axis) and standard deviation of the rank ($S_\mathtt{rank}$, y-axis)
of the correct reference when performing the identification procedure for 747 test
bacterial genomes, using 100 random reads and 3 times for each genome.
The closest the average rank is to 1 the closest to a perfect performance,
and the smallest the standard deviation of the ranks the least sensitive to sampling effects.
In order to increase clarity when many bacterial genomes tested produce equal or close coordinates on
the scatter, we use hexagonal binning and color the areas accordingly. The vertical bar on the right side of panel indicates the percentage
of times the correct reference was within the top 25 matches.
Various reads size (rows) and error rates (random substitution, columns) were tried, producing a matrix of scatter plots.}
\label{fig:perf_var}
\end{figure*}

For each bacterial genome, we took 100 random
simulated reads and scored them against a database comprising these bacterial genomes,
among a larger collection of sequences and genomes from other bacteria, phages, plant, fungi, viruses, 
and mammals using our method, recording the rank of the query genomes in
a list of the 25 best matching references (see Figure~\ref{fig:perf}). 
In order to assess the variability of the results for each test bacterial genome, this
was repeated 3 times for each genome. The
average ranks and the standard deviation for the ranks are presented in Figure~\ref{fig:perf_var}.

\begin{figure*}[!ht]
\begin{center}
\includegraphics[width=\textwidth]{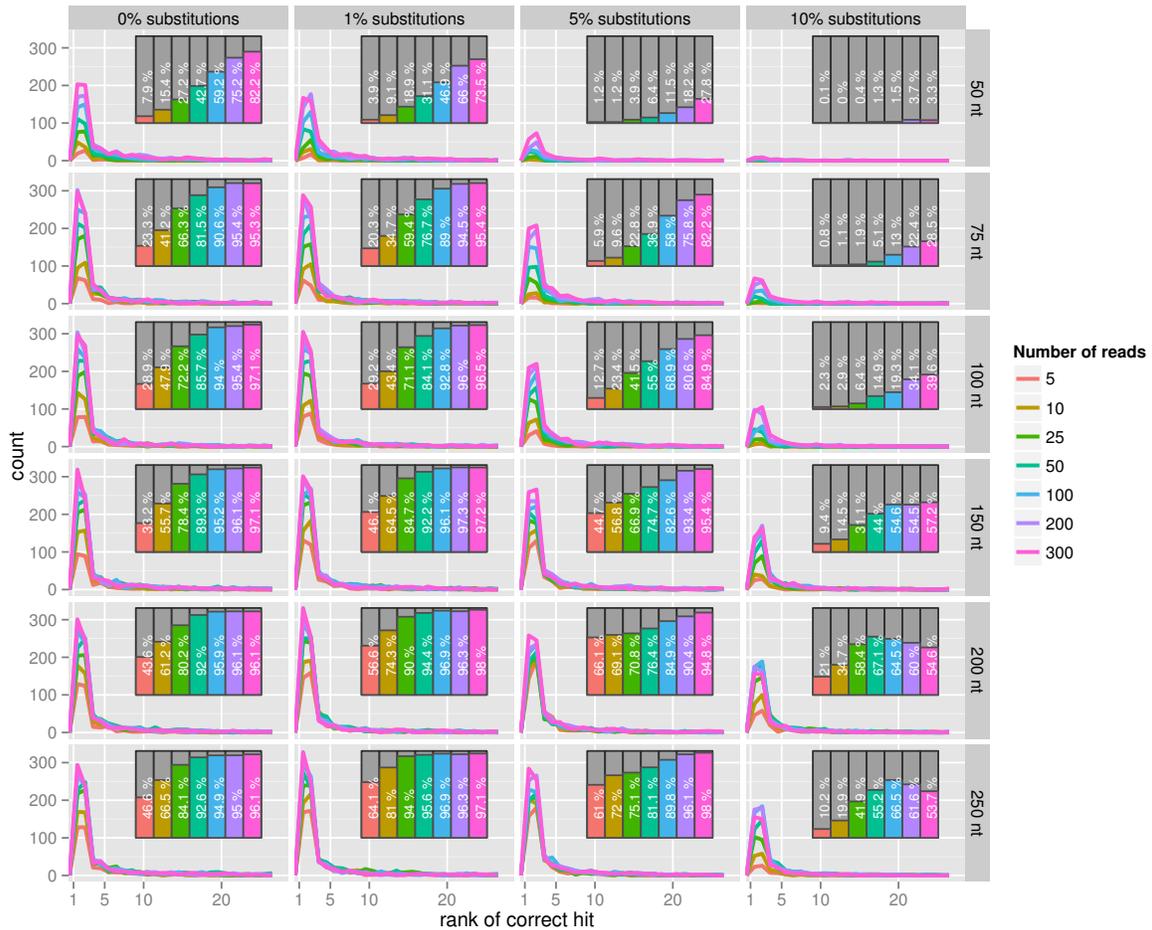}
\end{center}
\caption{{\bf Bacterial reads (number of reads).} 
For each bacterial genome in a set of 747 genomes, we simulated several read lengths
(50 nt, 75 nt, 100 nt, 150 nt, 200 nt, 250 nt) and several substitution error rates
(0\%, 1\%, 5\%, 10\%). Independent samples of 5, 10, 25, 50, 100, 200, or 300 random reads
were used in each query and the distribution
of the rank of the correct references in the list recorded; a rank of $1$ means that
the correct reference was at the very top of the list. 
The list of hits has a maximum length of 25 and we count the reference as `not found' if
it not present in the list. The percentages of correct test bacterial genomes found in that
list are represented in a bar plot nested on the right side of each panel.
Increasing the number of reads in the random sample beyond 100 reads
only improves very slightly the performances observed, mostly for shorter read lengths
and higher substitution rates. The substitution rate or the read length has much stronger effects
on the performances.
}
\label{fig:perf_nreads}
\end{figure*}

Performances were not satisfactory with reads of 50 bases in length, but we observed
dramatic improvements when increasing the read length, with reads of length 150 nt
already close to the maximum performances: the correct genome is then in the list of results, in the
top 5 hits over 97\% of 
the times with low substitution rates and in the top 15 with higher substitution rates (see Figure~\ref{fig:perf_nreads}).

Increasing the read length up to 250 bases helped partially compensate for the negative effect
of increasing substitution rates. Increasing the number of reads in the random sample
sent for identification could also compensate for shorter reads, or increasing substitution rates
up to $5\%$. Within the range of read lengths and number of reads we tested, a substitution error rate
of $10\%$ only permitted us to retrive the correct hit in less than $70\%$ of the cases. 

As detailed earlier our method aims at returning the right reference within a set of proposed
matches and by doing so simplifies the search space that a brute-force approach would require to
explore with computationally demanding procedures. Restricting ourselves to finding the query
sequence within the top five results is probably stricter than necessary, as running the analysis on all
25 would still be a significant improvement compared to an exhaustive search,
but emphasizes that the method
is already able to return the right answer within very small sets of candidate answers.

\begin{figure}[!ht]
\begin{center}
\includegraphics[width=.9\textwidth]{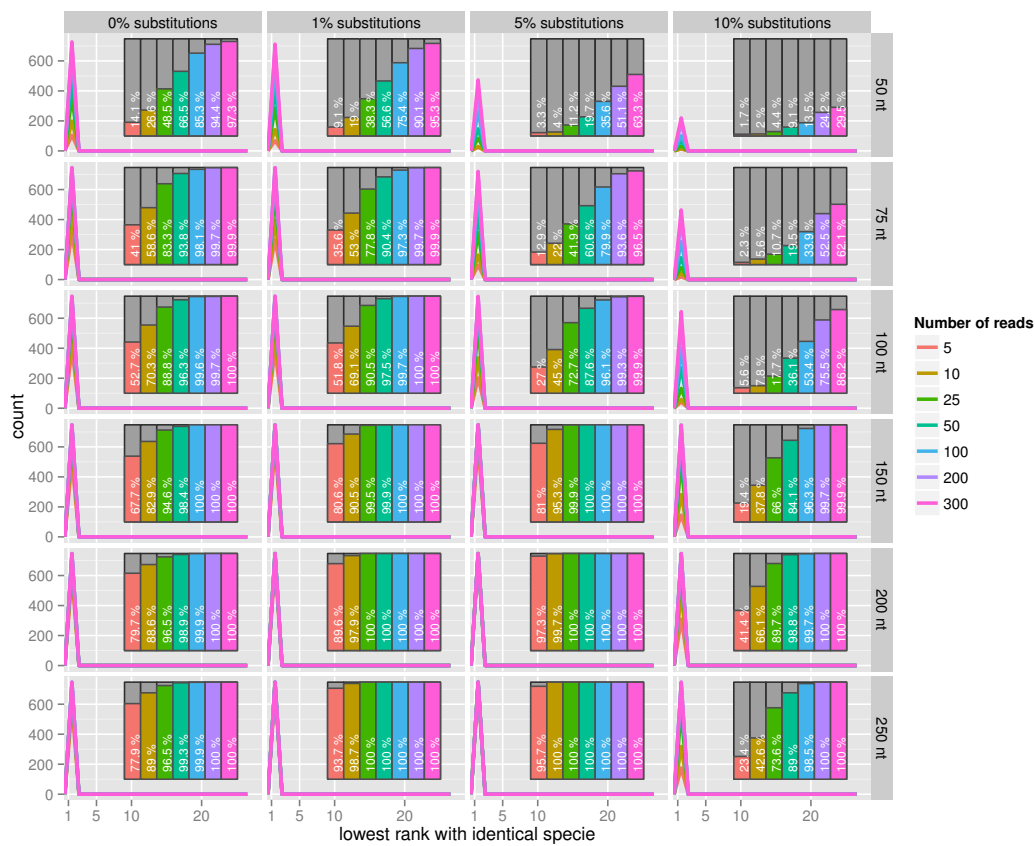}
\end{center}
\caption{{\bf Bacterial reads, same specie.}
Percentage of matches giving the correct specie, that is a reference in
our collection that belongs to a bacteria of the same specie 
rather than the correct exact same reference as shown in Figure~\ref{fig:perf},
and the percentage of cases for which the correct specie
was not in the top 25 matches. 
Independent samples of 5, 10, 25, 50, 100, 200, or 300 random reads
were used in each query and the distribution
of the rank of the correct references in the list recorded; a rank of $1$ means that
the correct reference was at the very top of the list. 
The list of hits has a maximum length of 25 and we count the reference as `not found' if
it not present in the list. The percentages of correct test bacterial genomes found in that
list are represented in a bar plot nested on the right side of each panel.
The performance remains poor for the shorter reads (50 nt), with noise decreasing it further (barplot on the first row), but become extremely good from 100 nt and
stays robust against noise.}
\label{fig:perf_species}
\end{figure}

In the context of iterative search and identification we consider that pointing out the right
bacterial specie, even if not the correct precise strain or genomic reference, is already a 
relatively successful answer.
Figure~\ref{fig:perf_species} shows that our identification procedure
performs relatively well, except with the shortest reads. It also shows that as little as
5 random reads of 200 nucleotides are enough to identify correctly the specie in up to over $90\%$
of the cases.

The range of lengths and substitution rates we used are comparable to the ones
obtained from next-generation sequencing platforms such as Illumina (maximum of 150 nucleotides with an error
rate of about 0.1-1\%, Life Technologies' SOLiD 5500 (maximum of 75 nt reads with an error rate of 0.01\%),
Ion Torrent PGM (maximum of 200-300 nucleotides with an error rate of 1\%), or Pacific Bioscience
(3,000 nucleotides with an error rate of 15\%). Our method performs well within these ranges,
and we anticipate to increase
performance further by adding support for paired-end sequencing.
Our method appears relatively insensitive
to sequencing errors such as base substitutions and the expected low ranks for our
test queries were minimally affected as substitution rates increased.

We have also tried the approach on sequencing data from Ion Torrent PGM from samples ranging from
viral and bacterial isolates to metagenomics mixtures. Very similar genomes in the 
collection of references indexed, such as several strains of the same specie, can contribute
to a degradation of the performances by increasing the probability of having closely related
genomes with lower ranks than the correct reference genomes. This is confirmed by
the increased performance when considering the species rather than the exact reference,
and this is a moderate inconvenience that can be disambiguated during a second iteration.
Finally, because we are considering {\it k}-mers within the context of reads rather than
breaking down all reads into {\it k}-mers and pooling all {\it k}-mers together for analysis,
we are obtaining very promising
results with sequencing from samples from diverse mammals, and anticipate to reliably identify
them in the near future.

\subsection*{Exhaustive computation on the sequencing reads}
The approach we present describes the use of a client-server approach for performing the analysis of sequencing
data without specifying a reference genome. The first component in the approach returns a list of hits
as candidate reference sequences, the guiding principle being to restrict the search space for more computationally
demanding methods, such as the alignment of all sequencing reads. We have shown that the correct reference
can be found in an ordered set of hits when using randomd sampling of reads on a client and a simple {\it k}-mers
based scoring algorithm on a server; as a proof-of-concept, we implemented the last step as an alignment
of all reads against the reference genomes returned in the list, and select the reference against which the largest
number of reads can be aligned.

  \subsection*{Computing performances}

     \subsubsection*{Server}
Memory usage on the server can be minimized by using a disk-based key value store, and tuning
performances can be achieved by caching this into the memory available on the computer running it. 
Thanks to the use of a NoSQL database, we also anticipate to be able to scale up as genomic data get
increasingly abundant, and continue being able to index and query increasingly large collections of
references on relatively affordable computer systems using sharding.

With the current implementation both the indexing system and the server are implemented in Python,
and the indexing of 44 Gbases of reference DNA is performed in a few hours using 8 cores
(Intel Xeon, 2.93GHz), and the processing
of one incoming sample taking under 10 seconds. We are aware that a significant speedup could be
achieved with optimization efforts such as bottlenecks moved to C, but it is also possible to increase
global performances in the handling of more requests by dedicating more cores, 
should the need arise.

     \subsubsection*{Clients}  
We have implemented a command-line tool to perform the complete workflow
in Figure~\ref{fig:workflow}.
The tool performs a random sampling of the sequencing reads, queries the server with these
reads, and retrieves a list of top hits. In a second step, it fetches the full reference genomes associated
with the reads and performs the alignment of all reads using bowtie2, keeping along each alignment
the unmapped reads. The alignment for which the lowest number of unmapped reads is obtained is
considered the best hit, and the umapped reads are sampled and queried in turn.
This is repeated until a specified maximum number of iterations is reached, there are no unmapped
reads remaining, or the list of candidate matches returned by the server is empty.

To facilitate the use of our method, and demonstrate the relatively low computational demand
of working with microbial genomes without the reference genome matching the sample being known,
we have also developed a browser-based client using Javascript
and HTML5 features that can be accessed at \url{http://tapir.cbs.dtu.dk}.
The client is currently only implementing the sampling and query,
and is generally working on the latest releases of Firefox and Chrome/Chromium.
Safari and Internet Explorer do not appear to support the HTML5 features
we require. Chromium v.25 running on a laptop with an Intel Core i7 CPU clocked at 2.10 GHz,
is able to process the raw reads in a 500 Mb FASTQ file in size could be processed in under
15 seconds, with few seconds spent communicating with the server.

\section*{Discussion}
The concept underlying TAPIR is rather simple, and to some extent a continuation of the
generic work around document search in computing and information technology, and of the
use of {\it k}-mers when working on DNA sequences\cite{ning_ssaha:_2001,zemin_ning_ssaha}.
The increase in size of DNA databases has
been announced and observed for at least a decade, but recent developments in DNA
sequencing technology have made fast and affordable generation of data a reality.
We argue that matching experimentally-obtained DNA sequences against all known
DNA is one of the most important challenges in bioinformatics. We show here that this can
be done with a speed and ease that matches what the internet web search giants have made us expect.

We also found the first step of the
approach to mapping proposed very recently by Liao and collaborators\cite{liao2013subread}
to be similar to the way we score hits for a given read, although both approaches were developed independently.
We plan on working further on both the implementation, moving the slowest parts to C in order to allow
an increased number of connections while keeping hardware requirements for the server relatively modest.

We have shown with synthetic sequencing reads from almost 750 bacteria that while our approach
does not always find the exact reference as the first hit, it can find it among the first hits in
over 97\% of the time, and can find the correct specie almost all the time. When coupled
with more detailed methods such as state-of-the-art aligners, we think of our approach as a way to
simplify the search space when confronted with unknown DNA, and the detailed methods have
to be performed only on that restricted number of candidate references. 

The results obtained from our general client-server approach also suggest that the analysis of NGS data could
benefit from taking naive approaches, in the sense that the expected reference sequence is not
provided, but determined by a query to a service such as TAPIR. We also observed that results obtained
with very few reads in the sample were better than anticipated for the longer read lengths,
and suggest that further work will be needed for assigning score for situations where only very and very few reads
are available. We implemented the computationally exhaustive step on the sequencing as an alignment of the reads
using bowtie2, and select the match for which the largest number of reads are aligned. This number is closely related
to the average coverage, and recent work shows that better measures can be found\cite{lindner2013analyzing}.
Further work will be required to refine the approach presented here.

When considering tasks such as real time surveillance, such as infections in patients, biodefense,
or food safety, modern desktop DNA sequencers such the Ion Torrent PGM or Illumina MiSeq are
already up to the task and our method provides an immediate early step during which
the search space can be narrowed down and more advanced analysis methods can be performed locally
afterwards, without the need to transfer large amounts of raw data between a laboratory performing the 
DNA sequencing and a computing facility.

Although initially designed for pure cultures, with the eventuality of a plasmid or a viral
component also present, we obtain very encouraging results with more complex samples with
reads from tens of different bacteria. We anticipate that this client-asks-server-for-hints
approach will develop more, with de-novo genome assembly methods even looking whether known
genomes can provide good templates and reserve the hardest computational task to the reads
that belong to genomes or genomic fragments not yet in a database such as ours.

Finally, our use of B-trees in NoSQL databases is less efficient than FM-Index searches
when perfect matches are considered. However, we argue that the semi-independence from the length
of the reference sequence FM-index search can achieve can be counterbalanced, at least in part,
by an increase in the number of different reference sequences. Such indexes must also be
contained into the memory of one computer, the only alternative being to build several indexes
from complementary subsets of reference sequences. Because of sampling, limited computing ressource
are needed, and B-trees or hashes can be split across
several compute nodes, a technique known as sharding, represents a larger benefit. 
\section*{Methods}

  \subsection*{Sources of genomic references}
Publicly available genomes, contigs, plasmids, and individual genes available from the 
EBI and the NCBI were downloaded to be our reference DNA. The exact composition of
the references will be expanding with time, but we listed the snapshot used for the
writing of the manuscript in Table~\ref{Tab:01}.

  \subsection*{Indexing of references}
Each reference sequence was split into non-overlapping {\it k}-mers and for
all {\it k}-mers across all references a key-value store, 
or NoSQL database (we used KyotoCabinet\cite{hirabayashi_kyoto}),
was created, associating to each k-mer (key in the database)
a list of identifiers corresponding to the references having that k-mer. 
We called this the presence database.
Similarly, the positions in the reference at which the k-mer is found were stored in what 
we call the position database.
$k$ was chosen to be equal to 16, as it gave us satisfactory results, and as a multiple of $4$
was well-suited for bit-packing.
The associations between references identifiers and information, such as a description line
and the source of data, were stored in a separate SQL database.

  \subsection*{Scoring}
In order to score a set of short query sequences, presumably sequencing reads, we iterate through a random
sample of them. The larger the sample size the more reliably accurate it will become.
For each read, we iterate over the consecutive {\it k}-mers obtained by sliding a window of
width {\it k} across the sequence.
For each {\it k}-mer within a read, if it has not been counted before and it is found in the presence database
we then query the position(s)
for the reference(s). Once all {\it k}-mers for a read have been processed, we look at the number
of approximate contiguous positions matched in the references and only consider the largest clusters
of matches, that is the largest concentration of matching {\it k}-mers originating from the same
read across all matching references. To achieve this, we are sorting the positions of the matching {\it k-mers}
on the reference and look at consecutive positions within a tolerance gap from one an other.
For each such cluster, we add the number of {\it k}-mers
to a possibly previously added number for that reference and we update the list of {\it k}-mers already
counted. The next sequence, or read, is then processed. When all reads have been processed
we obtain a list of references
to which is associated a count of matching {\it k}-mers. For each pair $<reference, count>$,
the count is divided by the number of unique {\it k}-mers in the query set, giving us a rough score
for the amount of DNA in the query matched by a given reference. If a query set completely
matches the sequence that score will be 1, it will be lower
 otherwise; for example, if the query set is a mixture in equal proportion of two references
the score would be around $0.5$ for both references.
That count is also divided by the size of the reference, giving a rough score
for the fraction of the reference that is represented by the query; that second score
is helpful to sort the matching references, and avoid bias toward the largest references.
The final score is a weighted sum of those two scores, default being equal weights.
If the query set is large, for example if we are considering all reads coming
out of a DNA sequencing run, we use only a random sample of that set.
The Python code for our server is available and can be consulted for full details about the implementation.

  \subsection*{Implementation of a client}
To facilitate the use of the service, we implemented an HTML5/Javascript client running as a page
in a web browser. Developement was made with Firefox 15 as a target platform, at it was the
only browser implementing all needed features on all platforms (Linux, Mac OS X, 
Microsoft Windows, as well as on Android 4.0 (tablet ASUS TF101 - we anticipate that it would 
also work from a high-end smartphone). At the time of writing, the latest release of Firefox
(v.21) is not working but the lastest release of Chrome (v.25) is working.

We have also made available a Python library and command-line tool for
easy integration in existing workflows and pipelines. This script is available as supplementary material
and we hope to expand it and release a Python package.

  \subsection*{Other technical specifications}
With the exception of bindings to libraries such as KyotoCabinet, all implementation was made using Python
version 2.7.3 on the server side. The web application uses the micro-framework Flask 
and is served by lighttp.
We have not yet dedicated much effort to optimization, such as moving specific parts of the code to the
C language to obtain gain in speed, as the performances are sufficient for our current needs.

The client-side library and command-line tool were developed for Python version 3.3.

\section*{Acknowledgments}

Aron Eklund for critical reading.

\bibliography{template}

\section*{Figure Legends}

\begin{figure}[!ht]
\begin{center}
\includegraphics[width=.9\textwidth]{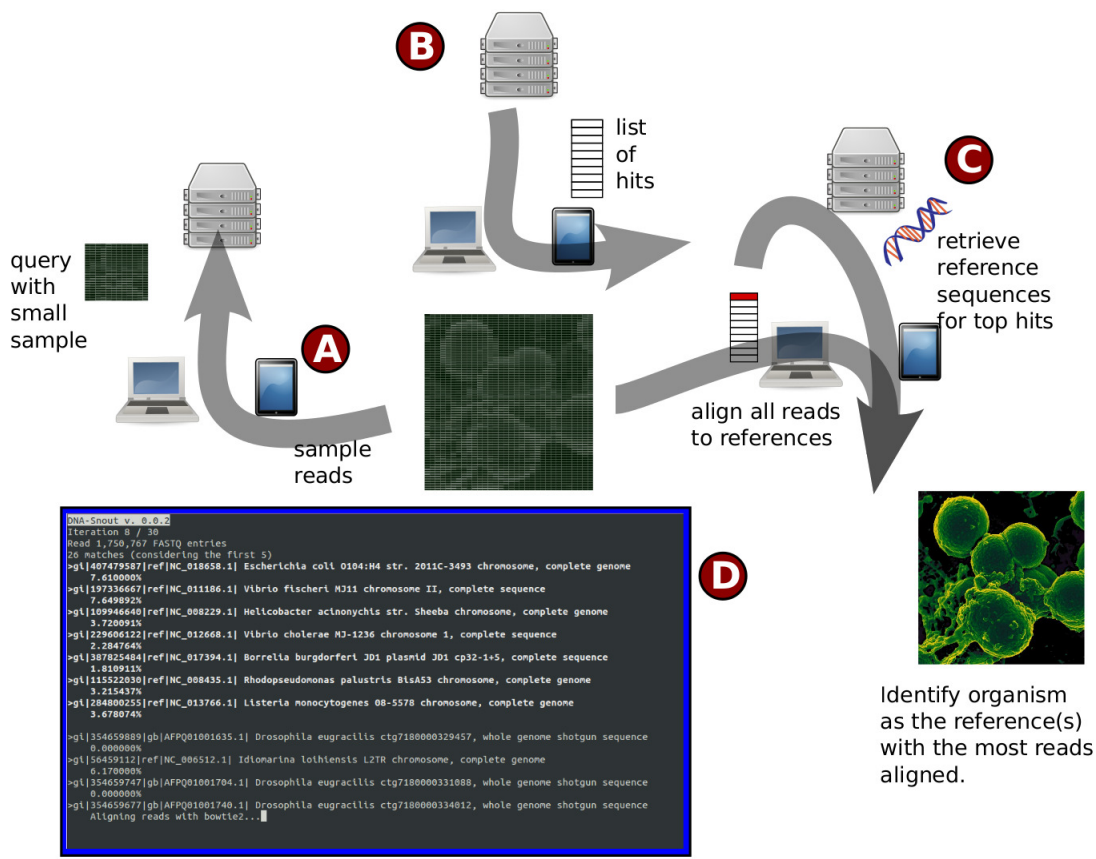}
\end{center}
\caption{{\bf Client-server alignment without pre-specified reference genome.}
(A) A small random sample of the umapped reads (initially all reads)
 is taken by the client and sent to the server. (B) In return
the server sends a list of hits, or candidate reference sequences for
the sample. (C) The client then iterates through the top hits and for each one requests
the full genomic sequence from the server after checking that it does not already have a copy of it locally,
and calls bowtie2 to build an index for that reference and align all currently unmapped reads to it.
The reference for which the most reads map is kept, and unmapped reads remaining are moved back
to step (A). The outcome is a list of reference genomes, along with a percentage of the reads iteratively
aligning to these references (screenshot in (D)).}
\label{fig:dnasnout-console}
\end{figure}


\section*{Tables}

\begin{table}[!ht]
  \caption{{\bf Genomic references}}
  \begin{tabular}{lrr}\hline
    & Number of references & Size (DNA bases) \\ \hline
	HIV&	4,053&	36,471,153 \\
	Phage genomes (EBI)&	1,078&	59,538,128\\
	Viral genomes (EBI)&	3,464&	64,859,892\\
	Bacterial genomes (EBI) &	747&	2,418,028,337\\
	Bacterial genes (NCBI)&	5,218,077&	4,963,568,551\\
	Bacterial genomes (NCBI)&	4,693&	8,584,324,670\\
	Viral genomes (NCBI)&	1,750&	60,637,755\\
	Fungi (NCBI)&	202,270&	298,736,207\\
	Human Microbiome sequences&	1,653,700&	1,490,442,185\\
	Plasmids (NCBI)&	159,705&	132,800,479\\
	Viruses (EBI)&	78,630&	65,110,952\\
	{\it Homo sapiens} (Hg19)&	3,134&	2,844,000,504\\
	{\it Mus musculus}&	305&	2,745,142,291\\
	Plant (RefSeq)&	558,267&	8,622,349,159\\
	Invertebrates (Genbank)&	1,123,813&	18,429,666,992\\
	Protozoa (Genbank)&	47,275&	1,997,449,553\\
	Fungi (Genbank)&	200&	242,402,709\\
        \hline
  \end{tabular}
  \begin{flushleft}
    Snapshot of genomic references (source and number of references).
    The references are a mixture
    of full genomes or plasmids, and of genomic fragments such as contigs or genes.
  \end{flushleft}
  \label{Tab:01}
\end{table}


\end{document}